\documentclass[11pt]{article}
\usepackage{amssymb,amsmath}
\begin{document}
\begin{titlepage}
\title{An Exact Cosmological Solution of the Coupled Einstein-Majorana
Fermion-Scalar Field Equations}
\author{A.~B. Balantekin\footnote{baha@physics.wisc.edu}\\
\small{  Department of Physics, University of Wisconsin}\\
              \small{ Madison, Wisconsin 53706 USA } \\ \\
T.~Dereli\footnote{ tdereli@ku.edu.tr}\\
\small{Department of Physics, Ko\c{c} University}\\
\small{34450 Sar{\i}yer, \.{I}stanbul, Turkey}}
\date{ }
\maketitle
\begin{abstract}
\noindent  We couple a neutral scalar field and a Majorana fermion
field to Einstein gravity represented by the Robertson-Walker
metric and find a class of exact cosmological solutions.
\end{abstract}

\vskip 1cm


\end{titlepage}

\section{Introduction}

Recent years witnessed progress at a breathtaking pace both in
cosmology and in neutrino physics. In cosmology, initial evidence
for the acceleration of the universe from high red-shift supernova
observations was subsequently verified by the precise measurements
of the cosmic microwave background radiation. These observations
and measurements discovered that there is a significant
"dark-energy" component of the universe. In a parallel development
solar, atmospheric, and reactor neutrino experiments firmly
established that neutrinos are massive and that they mix.
Mass-varying neutrinos were introduced in part as an attempt to
explain the origin of the cosmological dark-energy density and why
its magnitude is apparently coincidental with that of the measured
neutrino mass splittings \cite{Kaplan:2004dq,Fardon:2003eh}. Such
mass-varying neutrinos can behave as a negative pressure fluid
leading to the acceleration of the universe.

Models for the dark energy in which the energy density of the
scalar field approximates Einstein's cosmological constant were
studied in detail \cite{Farrar:2003uw}. It can be shown that in
the mass-varying neutrino scenario dark energy is also equivalent
to having a cosmological constant \cite{Peccei:2004sz}. It has
been argued that mass-varying neutrino models contain an
instability when neutrinos become non-relativistic and a stable
neutrino-varying mass model is indistinguishable from a
cosmological constant \cite{Afshordi:2005ym}. Mass-varying
neutrino models without acceleron like scalar fields were also
proposed \cite{Horvat:2005ua}. However, an analysis of the cosmic
microwave background radiation anisotropies and large scale
structure implies some evidence for coupling between neutrinos and
a scalar field \cite{Brookfield:2005bz}.

Quantization of a Fermi field, coupled to the Robertson-Walker
metric, had been worked out some time ago \cite{Isham:1974ci}.
Motivated by the recent studies of mass-varying neutrinos, in this
paper we explore if exact solutions exist when a neutral scalar
field (representing the acceleron of the mass-varying neutrino
models) as well as a Majorana fermion field (representing the
neutrino) couples to gravity via Robertson-Walker metric. We show
that such solutions indeed exist.

In the next section we outline our model. In Section 3 we present
our exact solution to this model. A brief discussion of our
results in Section 4 concludes the paper.

\section{The Cosmological Model}

We solve coupled Einstein-acceleron-Majorana neutrino field
equations derived  by a variational principle from the Lagrangian
density
\begin{equation}
{\cal{L}} = \frac{1}{2\kappa} {\cal{R}} *1 - \frac{1}{2} d\phi
\wedge *d\phi + V(\phi) *1 + i \bar{\psi} (\gamma \wedge *\nabla)
\psi  - i M(\phi) \bar{\psi} \psi *1
\end{equation}
where  $\psi$ is the 4-component (Majorana spinor) sterile
neutrino  field and $\phi$ is the (real,scalar) acceleron field;
$V(\phi)$ the acceleron potential function, $M(\phi)$  the
varying-mass of the neutral fermion to be determined. $\kappa = 8
\pi G$ is the gravitational coupling constant (in natural units
such that $c=1=\hbar$). The space-time geometry is given in terms
of a metric tensor $g$ of Lorentzian signature $-+++$ and its
unique Levi-Civita connection $\nabla$. ${\cal{R}}$ is the
corresponding curvature scalar and $*1$ denotes the oriented
volume element. We use a set of  real $\gamma$-matrices (Majorana
realization) $\{ \gamma_a\}$ that satisfy
\begin{equation}\gamma_a \gamma_b + \gamma_b \gamma_a = 2
\eta_{ab} I .\end{equation} Reality means ${{\gamma}^{*}}_a =
\gamma_a$. Hence we have ${\gamma^\dagger}_0 = {{\gamma}^T}_0 = -
\gamma_0$ and ${\gamma^\dagger}_i = {{\gamma}^T}_i = \gamma_i .$
The conjugate spinor field is \begin{equation} \bar{\psi} =
{\psi}^{\dagger}C, \end{equation} where the charge conjugation
matrix $C = \gamma_0 .$ With our conventions the spinor field
$\psi$ is self-conjugate if and only if it is real. That is
\begin{equation} \psi^C \equiv C {\bar{\psi}}^T
= \psi^* .\end{equation}

We look for exact solutions of the coupled system of field
equations in the Robertson-Walker space-time determined by the
metric (spatially flat, $k=0$)
\begin{equation}
g = - dt \otimes dt + R^2(t) ( dx \otimes dx + dy \otimes dy + dz
\otimes dz)
\end{equation}
in terms of the cosmic time $t$ and isotropic coordinates
$x^i:(x,y,z)$. $R(t) \geq 0$ is the expansion function. We further
adopt the ansatz
\begin{equation}\phi = \phi(t)\end{equation} and
\begin{equation} \psi = ( h_1(t) + h_2(t) \gamma_0 )\xi
\quad , \quad  \bar{\psi}  =  \bar{\xi} (h_1(t) - h_2(t) \gamma_0)
\end{equation}
where $h_1(t) , h_2(t)$ are functions to be determined and $\xi$
is a constant Majorana  4-spinor. We substitute these in the
variational field equations and reduce them to a set of ordinary
differential equations:
\begin{eqnarray}
\frac{3}{\kappa}\left( \frac{\dot{R}}{R} \right)^2 -
\frac{{\dot{\phi}}^2}{2} + V(\phi) &=& \mathcal{N} M(\phi)
({h_1}^2 + {h_2}^2)
\\
\frac{2}{\kappa}\frac{\ddot{R}}{R}+\frac{1}{\kappa} \left(
\frac{\dot{R}}{R} \right)^2+\frac{{\dot{\phi}}^2}{2}+
V(\phi)&=&\mathcal{N} (h_2\dot{h}_1-h_1\dot{h}_2) \nonumber
\\ & & + \mathcal{N} M(\phi) ({h_1}^2 + {h_2}^2)
 \\
\ddot{\phi} + 3 \frac{\dot{R}}{R} \dot{\phi} -  \frac{dV}{d\phi}
&=& - \mathcal{N} (\frac{dM}{d\phi})({h_1}^2+ {h_2}^2)
 \\
\dot{h}_1 + \frac{3\dot{R}}{2R} h_1 + M(\phi) h_2 &=& 0
\\
\dot{h}_2 + \frac{3\dot{R}}{2R} h_2 -  M(\phi) h_1 &=& 0
\end{eqnarray}
where
\begin{equation} i(\bar{\xi} \xi) \equiv \mathcal{N} \end{equation}
may be (semi-classically) interpreted  as a (real) vacuum
expectation value.

\noindent {\bf Remark: 1} If we differentiate (8) and simplify by
using (10), (11), (12)  and (8) again, we obtain precisely (9).
Thus, out of the five equations above,  only four of them are
independent. On the other hand we have four functions to solve
for. Therefore the system is well-determined.

\noindent {\bf Remark: 2} If we use (11) and (12) for the
derivatives of $h_1$ and $h_2$ and substitute in (9), the right
hand side vanishes. This means that in our simple model the
neutrino pressure is zero.

\noindent {\bf Remark: 3}  (11) and (12) implies that
\begin{equation} R^3 ({h_1}^2 + {h_2}^2) = C \end{equation}
is a constant of motion.  We designate $ n = \mathcal{N }C  $
which may be called the neutrino number density.

\section{An Exact Solution}

We only consider the simple case of a constant potential. $V(\phi)
= V_0$. First we solve the resulting equations,
\begin{eqnarray}
\frac{2}{\kappa} \frac{\ddot{R}}{R} + \frac{1}{\kappa} \left(
\frac{\dot{R}}{R} \right)^2 + \frac{{\dot{\phi}}^2}{2} + V_0 &=&
0, \label{a1}
 \\
\frac{3}{\kappa} \left( \frac{\dot{R}}{R} \right)^2 -
\frac{{\dot{\phi}}^2}{2} + V_0 - \frac{n}{R^3} M(\phi) &=& 0,
\label{a2} \\
 \ddot{\phi} + 3 \frac{\dot{R}}{R} \dot{\phi} + \frac{n}{R^3}
 \frac{dM}{d\phi} &=&
0, \label{a3}
\end{eqnarray}
by assuming the existence of solutions that satisfy
\begin{equation} R(t) = R_0 e^{\alpha \phi(t)} \end{equation}
where both $ R_0$ and $\alpha$ are constants that may be chosen
later. We then solve (15) and get
\begin{equation}
\dot{\phi}(t) = \sqrt{\frac{2\kappa V_0}{\kappa + 6 \alpha^2}}
\cot \left( \sqrt{\frac{\kappa V_0}{8 \alpha^2} (\kappa + 6
\alpha^2)} t \right) .
\end{equation}
Inserting this expression into (16) determines the functional form
of $M(\phi)$. In order to get a closed expression for it, we first
write $\phi$ as a function of $t$:
\begin{equation}
\phi(t) = \phi_0 - \frac{2\alpha}{\kappa + 6 \alpha^2} \ln \left |
\csc^2 \left( \sqrt{\frac{\kappa V_0}{8 \alpha^2} (\kappa + 6
\alpha^2)} t \right) \right | .\end{equation} Here $\phi_0$ is an
integration constant. Taking the exponential of both sides and
using trigonometric identities we find
\begin{equation}\cot^2 \left( \sqrt{\frac{\kappa V_0}{8 \alpha^2}
(\kappa + 6 \alpha^2)} t \right) = e^{- \frac{\kappa + 6
\alpha^2}{2\alpha}(\phi - \phi_0)} - 1 .\end{equation}
Substituting this in (16), we solve for the mass function as a
function of $\phi$:
\begin{equation}
n M(\phi) = \frac{2\kappa V_0 {R_0}^3}{\kappa + 6 \alpha^2}
\left\{ e^{3\alpha \phi} +\frac{6 \alpha^2 - \kappa}{2\kappa}
e^{\frac{\kappa + 6 \alpha^2}{2\alpha}\phi_0} e^{-
\frac{\kappa}{2\alpha} \phi} \right\} .
\end{equation}
The mass-function may be given a better parametrization in terms
of its critical value such that
\begin{equation} \left. \frac{dM}{d\phi}\right |_{\phi = \phi_c} = 0 .
\end{equation}
The critical value, $\phi_C$, may be determined from the relation
\begin{equation}
e^{\frac{\kappa + 6 \alpha^2}{2\alpha}\phi_0} = \frac{12
\alpha^2}{6\alpha^2 - \kappa}  e^{\frac{\kappa + 6
\alpha^2}{2\alpha}\phi_C}.
\end{equation}
Therefore we may write
\begin{equation}
n M(\phi) = \frac{2\kappa V_0 {R_0}^3}{\kappa + 6 \alpha^2}
e^{3\alpha \phi_C} \left\{  e^{3\alpha (\phi - \phi_C)} + \frac{6
\alpha^2}{\kappa}e^{-\frac{\kappa}{2\alpha} (\phi - \phi_C)}
\right\} .
\end{equation}
The minimum mass value is
\begin{equation}
n M(\phi_C) = 2 V_0 {R_0}^3 e^{3\alpha\phi_C} .
\end{equation}

Differentiating with respect to $\phi$  we find
\begin{equation}
n \frac{dM}{d\phi} e^{-3\alpha \phi} = \frac{6\alpha \kappa V_0
{R_0}^3}{\kappa + 6\alpha^2} \left\{ 1 - e^{-\frac{\kappa + 6
\alpha^2}{2\alpha}(\phi -\phi_C)} \right\}.
\end{equation}
It is now possible to substitute this expression into (17) and
verify that it is identically satisfied. This is a consistency
check.

To complete the solution, we go to the Dirac system:
\begin{eqnarray}
\dot{h}_1 + \frac{3\dot{R}}{2R} h_1 + M(\phi) h_2 &=& 0 ,
\nonumber \\
\dot{h}_2 + \frac{3\dot{R}}{2R} h_2 -  M(\phi) h_1 &=& 0 .
\nonumber
\end{eqnarray}
Defining $ X(t) = e^{\frac{3}{2} \alpha \phi} h_1(t)$ and $Y(t) =
e^{\frac{3}{2} \alpha \phi} h_2(t)$ the above system can be
written in matrix form:
\begin{equation}
\frac{d}{dt}  \left ( \begin{array}{c} X(t) \\ Y(t) \end{array}
\right ) = \left ( \begin{array}{cc} 0&-M(\phi)  \\ M(\phi)&0
\end{array} \right )\left ( \begin{array}{c} X(t) \\ Y(t) \end{array}
\right ) .
\end{equation}
This can be integrated formally and the general solution reads
\begin{eqnarray}
h_1(t) &=& e^{-\frac{3}{2} \alpha \phi} \{ A \cos \Omega(t) + B
\sin \Omega(t) \}\nonumber \\
h_2(t) &=& e^{-\frac{3}{2} \alpha \phi} \{ A \sin \Omega(t) - B
\cos \Omega(t)  \} \nonumber
\end{eqnarray}
where $A = X(0) , B = Y(0)$ are the initial values and
\begin{equation}
\Omega(t) = \int^t_0 M(t') dt' .
\end{equation}
The mass is a complicated function of $t$, so we will leave the
solution implicit in terms of $\Omega(t)$. In fact the Majorana
spinor field for the solution is of the form
\begin{equation}
\psi = e^{-\frac{3}{2}\alpha \phi + \gamma_0 \Omega} (A -
B\gamma_o) \xi .\end{equation}

To conclude, we work out the expansion function as a function of
$t$
\begin{equation} R(t) = R_0 e^{\alpha \phi_C}
\left | \frac{12 \alpha^2}{6\alpha^2 - \kappa} \sin^2 \left(
\sqrt{\frac{\kappa V_0}{8 \alpha^2}(\kappa + 6 \alpha^2)}t \right)
\right |^{\frac{2\alpha^2}{\kappa + 6\alpha^2}} .
\end{equation}
It is not difficult to determine the following observable
quantities; the Hubble parameter:
\begin{equation}
H(t) \equiv \frac{\dot{R}}{R} = \sqrt{\frac{2\kappa\alpha^2
V_0}{\kappa + 6\alpha^2}} \cot \left( \sqrt{\frac{\kappa V_0}{8
\alpha^2}(\kappa + 6 \alpha^2)}t \right) ,
\end{equation}
and the acceleration parameter:
\begin{equation}
\frac{R\ddot{R}}{{\dot{R}}^2} = 1 - \frac{\kappa +
6\alpha^2}{4\alpha^2} \csc^2 \left( \sqrt{\frac{\kappa V_0}{8
\alpha^2}(\kappa + 6 \alpha^2)}t \right) .
\end{equation}

In the limit where $\kappa \ll \alpha^2$ and $V_0 \sim 1$,  to the
first order of approximation we would have
\begin{equation} R(t) \sim t^{\frac{2}{3}},
 \end{equation}
i.e. a matter-dominated Friedmann universe.

We next consider the case of a cosmological constant: $V_0 = -
\Lambda$. In this case Eqs. (\ref{a1}), (\ref{a2}) and (\ref{a3})
can be considered representing a generalization of the
$\Lambda_{\rm CDM}$ model \cite{Gron:2002}. The solution of Eq.
(\ref{a1}) now becomes
\begin{equation}
\dot{\phi}(t) = \sqrt{\frac{2\kappa \Lambda}{\kappa + 6 \alpha^2}}
\coth \left( \sqrt{\frac{\kappa \Lambda}{8 \alpha^2} (\kappa + 6
\alpha^2)} t \right) ,
\end{equation}
yielding
\begin{equation}
\phi(t) = \phi_0 + \frac{2\alpha}{\kappa + 6 \alpha^2} \ln \left |
\sinh^2 \left( \sqrt{\frac{\kappa \Lambda}{8 \alpha^2} (\kappa + 6
\alpha^2)} t \right) \right | ,
\end{equation}
and
\begin{equation}
\coth^2 \left( \sqrt{\frac{\kappa \Lambda}{8 \alpha^2} (\kappa + 6
\alpha^2)} t \right) = e^{- \frac{\kappa + 6
\alpha^2}{2\alpha}(\phi - \phi_0)} + 1 .
\end{equation}
In this case we find the mass function to be
\begin{equation}
n M(\phi) = \frac{2\kappa \Lambda {R_0}^3}{\kappa + 6 \alpha^2}
\left\{ - e^{3\alpha \phi} +\frac{6 \alpha^2 - \kappa}{2\kappa}
e^{\frac{\kappa + 6 \alpha^2}{2\alpha}\phi_0} e^{-
\frac{\kappa}{2\alpha} \phi} \right\} .
\end{equation}
Rest of the solutions of the Dirac system is unchanged. The
expansion function now has the form
\begin{equation}
\label{aa1} R(t) = R_0 e^{\alpha \phi_0} \left[ \sinh \left(
\sqrt{\frac{\kappa \Lambda}{8 \alpha^2}(\kappa + 6 \alpha^2)}t
\right) \right]^{\frac{4\alpha^2}{\kappa + 6\alpha^2}} .
\end{equation}
We also note the forms of the Hubble parameter
\begin{equation}
H(t) = \sqrt{\frac{2\kappa\alpha^2 \Lambda}{\kappa + 6\alpha^2}}
\coth \left( \sqrt{\frac{\kappa \Lambda}{8 \alpha^2}(\kappa + 6
\alpha^2)}t \right) ,
\end{equation}
and the acceleration parameter:
\begin{equation}
\frac{R\ddot{R}}{{\dot{R}}^2} = 1 - \frac{\kappa +
6\alpha^2}{4\alpha^2} \cosh^{-2} \left( \sqrt{\frac{\kappa
\Lambda}{8 \alpha^2}(\kappa + 6 \alpha^2)}t \right) .
\end{equation}
In the limit $\kappa \ll \alpha^2$ and $\Lambda \sim 1$, Eq.
(\ref{aa1}) again yields the matter-dominated Friedmann universe,
$R(t) \sim t^{2/3}$. In this limit ( for $t < t_\Lambda $) we get
the Hubble parameter to be
\begin{equation}
H(t) \sim \frac{1}{t_{\Lambda}} \coth (t/t_{\Lambda}),
\end{equation}
and the expansion factor to be
\begin{equation}
R(t) \sim \sinh^{2/3} \left( \frac{t}{t_{\Lambda}} \right),
\end{equation}
where $t_{\Lambda}^{-1} = (\sqrt{ 3 \kappa \Lambda})/2$. These
values of the Hubble parameter and the expansion factor are those
obtained in the flat $\Lambda_{\rm CDM}$ model \cite{Gron:2002}.
Hence the present model can be considered as a generalization of
the $\Lambda_{\rm CDM}$ model.

\section{Conclusions}

In this paper we found exact solutions to a cosmological model
where a neutral scalar field and a Majorana fermion field are
coupled to gravity represented by the Robertson-Walker metric. Our
study was motivated by the mass-varying neutrino models where the
 scalar field could represent the acceleron field and the
Majorana fermion field the neutrino. However, our results are
valid in a broader context, generalizing the earlier solution of
Isham and Nelson. The exact solution we found has a number of
parameters: the constants $R_0$ and $\alpha$ of the expansion
function and the critical value of the scalar field, $\phi_c$.
These parameters can be adjusted to construct different
phenomenological models.

\section*{Acknowledgement}

\noindent This  work   was supported in  part  by   the  U.S.
National Science Foundation Grants No. INT-0352192 and PHY-0555231
at the University of  Wisconsin, and  in  part by  the University
of Wisconsin Research Committee   with  funds  granted by the
Wisconsin Alumni  Research Foundation. We also acknowledge support
through TUBITAK-NSF Joint Research Project TBAG-U/84(103T113). We
thank the referee for helpful remarks and for bringing
Ref.\cite{Gron:2002} to our attention.


\end{document}